# Safeguarding the unseen: a study on data privacy in DeFi protocols

**Zhuangtong Huang, Jiawei Zhu, Zhongyu Huang, Yixin Xu, Jerome Yen, Ye Wang***

University of Macau, Macau, China

* Correspondence author; E-mail: wangye@um.edu.mo.

**Abstract:** The financial sector's adoption of technology-driven data analysis has enhanced operational efficiency and revenue generation by leveraging personal sensitive data. However, the inherent characteristics of blockchain hinder decentralized finance (DeFi) from accessing necessary sensitive user data. To address this challenge, we introduce a protocol that both safeguards user privacy and ensures data availability through the incorporation of homomorphic encryption and zero-knowledge-proof techniques in blockchain technology. This novel protocol helps mitigate privacy risks caused by sensitive data leaks while improving the capital efficiency of the DeFi market. Furthermore, we explore the applicability of these privacy-preserving methods in on-chain ecosystems and cross-border financial applications. Our solution contributes to secure, user-centric solutions for DeFi while upholding principles of decentralization and privacy protection.

## 1. Introduction

The ongoing digital transformation, especially in data-intensive sectors like the financial market, has given rise to a significant trend in the financial industry: technology-driven data analysis. This trend empowers financial institutions to harness substantial volumes of personal data to enhance operational efficiencies and maximize revenue [1]. However, this heightened data utilization has, in turn, engendered concerns due to the absence of robust privacy protection mechanisms within financial institutions, thereby resulting in more data breaches occurring, inflicting substantial financial losses. In response, nations have instituted specific data protection laws to address these critical issues [2]. Consequently, the challenge of balancing the need for financial institutions to utilize private data for enhancing financial efficiency and the equally crucial need for users to safeguard their personal privacy data merits heightened attention in the social discourse.

This challenge is particularly pronounced in the context of Decentralized Finance (DeFi) due to the inherent characteristics of blockchain technology [3]. DeFi is a rapidly expanding financial service domain characterized by financial applications executed through smart contracts within the blockchain infrastructure. The DeFi market has exhibited remarkable growth, with its size projected to exceed 100 billion USD by 2022 [4]. However, this rapid expansion has introduced unique challenges related to utilizing user privacy data to enhance market efficiency while safeguarding user privacy [5]. The transparency of the blockchain within the DeFi landscape poses a distinctive problem [6]. When sensitive information is uploaded onto the blockchain, it becomes publicly accessible, which discourages users from sharing their sensitive data despite the potential benefits of tailored services. Hence, this limitation also prevents DeFi protocol from utilizing multiple sources of sensitive data to design financial services that cater to a broader user base and improve efficiency [7]. Addressing this requires a protocol that can integrate various data sources' accessibility while preserving user privacy to meet the needs of both DeFi protocols and users.

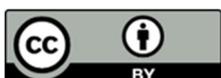

Previous studies have employed various techniques to tackle privacy challenges in blockchain applications [8,9], such as the Internet of Things [10] and electronic voting [11]. Nevertheless, within a DeFi framework, these solutions cannot provide both the accessibility of multiple sensitive data sources and the protection of user privacy [6]. A particularly glaring shortcoming of these solutions is that, in the event of a data breach, they risk exposing all private data, potentially leading to significant losses for users. In this paper, we address these two challenges by introducing a protocol that allows DeFi protocols to process user data on public blockchains without revealing essential details, including user identity and asset particulars. Our proposed system encompasses three primary components: trusted sources, zero-knowledge proof service provider (ZKPSP), and relayer. It capitalizes on asymmetric encryption, homomorphic encryption, and zero-knowledge proofs to ensure data confidentiality and accessibility.

The system operates in two distinct phases, which are the asset uploading phase and the asset verification phase. During the asset uploading phase, DeFi users interact with trusted sources and the relayer to upload their sensitive data to the blockchain system. In the asset verification phase, DeFi protocols can communicate with the ZKPSP to obtain a verification of users' assets to determine whether to proceed with the user's request. Throughout the entire information processing stage, no entity can match the user's identity or exact asset data, ensuring complete data privacy protection. Additionally, the DeFi protocol only obtains minimal personal private data, sufficient to enhance market efficiency.

Our contribution is distinguished in two primary areas. Firstly, we ensure that users' private data remains secure on the public chain. Notably, no single entity within our system can access all of a user's private data, significantly reducing privacy risks even in the event of data breaches. Secondly, we prioritize data availability while maintaining privacy, allowing DeFi protocols to enhance services using private details like user asset information. This addresses the longstanding issue of prior DeFi protocols' limitations in accessing sensitive user data. Consequently, our approach promises to advance the DeFi landscape, even further paving the way for novel protocols that adeptly integrate private user data from both on-chain and off-chain origins.

## 2. Related work

In this section, we present an overview of prior research in the domain of blockchain security and privacy preservation measures. Earlier studies predominantly concentrated on addressing two fundamental aspects of privacy within blockchain systems [12,13]: firstly, transactional unlinkability privacy, which conceals or obfuscates the connections between transactions to prevent their visibility or discovery; and secondly, content confidentiality privacy, which ensures that transaction contents remain exclusive to their involved parties.

In the first effort to protect the transactional unlinkability privacy, prominent techniques in prior work include mixing services, ring signatures, and Zero-knowledge proof (ZKP). Mixing services, introduced initially by Chaum [14], allow users to obscure both the identity of communication participants and the content of the communication, serving as a means to obfuscate transaction histories and reduce the risk of de-anonymization when integrated into blockchain networks. Ring signatures enable a user, as a part of a 'ring' of members, to sign a message without revealing the specific member responsible for the signature [15]. ZKP is a cryptographic technique designed to establish the validity of a given statement without revealing any additional information [16]. Researchers have explored enhancing anonymity in blockchain by involving those technologies [17]. Mahmood *et al.* integrated ring signatures and Zero-Knowledge Proofs (ZKPs) to conceal users' wallet addresses [18]. Some researchers ensure the transfer of closed data without the loss of reliability and with respect for the privacy of group members based on blockchain and linkable ring signatures [9,11]. Zerocoin, a distributed e-cash system, utilized Non-Interactive Zero-Knowledge (NIZK) proof cryptography to break the link between individual Bitcoin transactions [19], and the research further introduced significant enhancements, resulting as Zerocash [20]. Wang *et al.* proposed an electronic voting scheme based on blockchain, catering to large-scale voting, and combining homomorphic encryption with ring signatures [21]. However, these approaches primarily address the issue of transaction unlinkability and do not adequately tackle the challenge of safeguarding sensitive data when utilized in DeFi scenarios.

In the second effort to protect content confidentiality privacy, the homomorphic cryptosystem serves as a pivotal approach. A homomorphic cryptosystem (HC) is such a cryptographic encryption methodology that satisfies homomorphism so as to preserve arithmetic operations carried out on ciphertexts [22]. Prior researchers have affirmed the effectiveness of homomorphic encryption algorithms in conjunction with blockchain technology for private data protection, employing methodologies such as simulation experiments [23,24]. Consequently, some privacy protection systems merging homomorphic encryption and blockchain have emerged to address a wide array of application scenarios, spanning circuit copyright protection contexts [25], deployment in edge computing [23], utilization in smart grids [24], and implementation within IoT systems [10].

In addition to HCs, there have been other notable contributions towards content confidentiality and privacy. Pauwels *et al.* introduced the zkKYC system, adeptly balancing user privacy with regulatory demands for anti-money laundering (AML) and know-your-customer (KYC) [26]. Zhang *et al.* proposed a novel approach using decentralized oracles (DECO) which assures transaction outcome verification without divulging transaction details through zero-knowledge proof [27]. Hawk presents a decentralized smart contract system focusing on data protection during on-chain activities [28]. ZEXE, outlined by Bowe *et al.*, is a ledger-based framework that prioritizes both data and function privacy. It enables users to conduct offline computations to ensure privacy before initiating online transactions, leveraging the decentralized private computation (DPC) model [29].

However, it's worth noting that these solutions do not fully align with the requirements of sensitive data utilization in DeFi. Some solutions failed to meet the privacy protection needs due to certain solutions having entities capable of aggregating all data sources, which poses a substantial risk if compromised [23,24,26,28]. Moreover, several solutions are inherently designed for private blockchain or consortium blockchain, while DeFi predominantly operates on public chains [10,29]. Other challenges include the absence of necessary data aggregation functionalities [27], computational capabilities [25], and data validation functions [23] critical for DeFi data availability implementations.

Table 1 presents a comparison of blockchain-based privacy protection solutions. In DeFi, personalized services necessitate integrating, encrypting, and validating diverse data on public blockchains [30]. Given the inherent sensitivity of user data coupled with adversarial threats in DeFi [31], there is a pressing need for enhanced robustness against potential breaches. Current solutions tend to neglect aspects like specific privacy [18,21,19], suitability for public blockchains [10,29], comprehensive privacy robustness [23,24,26,28], or data availability [23,27]. Our solution aims to address these gaps by introducing a framework that combines ZKP, homomorphic encryption, and blockchain, tailored for DeFi's specific privacy requirements. This integrated solution, on the one hand, combines encryption, computation, and verification capabilities, enhancing data accessibility for DeFi protocols. On the other hand, our solution ensures such robustness that no participant can decipher the precise sensitive data of individual users, which achieves both privacy preservation and data availability.

**Table 1.** Comparison of privacy protection solutions on the blockchain. Abbreviations: App. (Application), Privacy (Protected Privacy), Public Chains (Applicable to Public Chains), Comp. Func. (Computational Function), Data Val. Func. (Data Validation Function).

| Authors | App. | Privacy | Public Chains | Comp. Func. | Data Val. Func. |
|---|---|---|---|---|---|
| Mahmood *et al.* [18] | GDPR-compliant blockchain | Transactional unlinkability | No, consortium blockchain | Not involving | Not involving |
| Wang *et al.* [21] | E-voting scheme | Transactional unlinkability | Yes | Not involving | Yes |
| Liu *et al.* [17] | Coin mixing | Transactional unlinkability | Yes | Not involving | Not involving |
| Miers *et al.* [19] | E-cash | Transactional unlinkability | Yes | Not involving | Yes |
| Sasson *et al.* [20] | E-cash | Transactional unlinkability | Yes | Not involving | Yes |

**Table 1.** *Cont.*

| Authors | App. | Privacy | Public Chains | Comp. Func. | Data Val. Func. |
|---|---|---|---|---|---|
| Liang *et al.* [25] | Circuit copyright | Content confidentiality | Not involving | Not involving | Not involving |
| ukil *et al.* [10] | IoT | Content confidentiality | No, private blockchain | Not involving | Not involving |
| Singh *et al.* [24] | Smart grid | Content confidentiality | Not involving | Yes | Not involving |
| Yan *et al.* [23] | Edge computing | Content confidentiality | Not involving | Yes | Not involving |
| Pauwels *et al.* [26] | AML and KYC system | Content confidentiality | Yes | Not involving | Yes |
| Zhang *et al.* [27] | Decentralized oracles | Content confidentiality | Yes | Not involving | Yes |
| Bowe *et al.* [29] | Ledger-based system | Content confidentiality | Not involving | Yes | Yes |
| Kosba *et al.* [28] | Decentralized smart contract system | Content confidentiality | Yes | Yes | Not involving |
| Our work | DeFi | Content confidentiality | Yes | Yes | Yes |

## 3. Background on cryptographic primitives

In this section, we present the essential background knowledge of various cryptographic primitives that form the foundation of our proposed solution. Cryptographic primitives play a critical role in ensuring the security and privacy of the proposed system. Our solution builds upon established cryptographic techniques, including homomorphic encryption, ZKP, and asymmetric key encryption schemes. We provide a brief overview of these primitives and their applications in the context of our proposed solution. An understanding of these cryptographic primitives is essential to grasp the technical details of our proposed system.

*3.1. Asymmetric key encryption*

Asymmetric encryption schemes are a fundamental tool in modern cryptography. They are based on a pair of keys, namely the public key *pk* and the private key *sk*. The encryption function *Encrypt* () takes as input a plaintext message m and the public key *pk*, and returns a ciphertext *C*. The decryption function *Decrypt* () takes as input the ciphertext *C* and the private key *sk*, and returns the original plaintext message *m*.

The public key *pk* can be shared with anyone in the system, while the private key *sk* is kept secret by the owner of the encryption scheme. This allows for secure communication between two parties without the need for a shared secret.

The RSA algorithm is one of the most well-known asymmetric encryption algorithms. It was introduced in 1978 by Rivest, Shamir, and Adleman [32] and is widely used in practice. The security of RSA is based on the difficulty of factoring large integers, which is believed to be a hard problem for classical computers [33].

*3.2. Homomorphic encryption*

Homomorphic encryption has gained significant attention as a promising cryptographic technique to enable secure computation with encrypted data. The technique allows for algebraic computations to be performed on ciphertexts, and the decrypted result is identical to the output that would be obtained if the computation was performed on plaintext data.

A homomorphic encryption scheme is comprised of a public key *pk* and a private key *sk*, an encryption function *Encrypt* (), a decryption function *Decrypt* (), and two computation functions: the original function $f$ (), and the homomorphic encryption function $Hom_f$.

Suppose that $m_1$ and $m_2$ represent the plaintext messages, and $f(m_1, m_2)$ is the result of a computation performed on these plaintext messages. The encryption function *Encrypt* () is used to encrypt the plaintexts, resulting in the ciphertexts $C_1$ and $C_2$:

$C_1 = Encrypt\,(m_1, pk)$, $C_2 = Encrypt\,(m_2, pk)$.

The homomorphic encryption function $Hom_f$ is used to perform algebraic computation on the ciphertexts $C_1$ and $C_2$, resulting in the ciphertext $C_3$:

$C_3 = Hom_f\,(C_1, C_2, pk)$.

The homomorphic encryption scheme ensures that the computation results of the plaintexts $m_1$ and $m_2$ can be decrypted from the computation results of the ciphertexts $C_1$ and $C_2$ using the decryption function *Decrypt* ():

$Decrypt\,(C_3, sk) = f(m_1, m_2)$.

The concept of homomorphic encryption was first proposed by Rivest *et al.* [32]. Previous studies have shown that homomorphic encryption algorithms are highly resistant to Chosen Ciphertext Attacks (CCA) [34]. Therefore, homomorphic encryption has practical applications in scenarios where privacy-sensitive and computation-intensive tasks are required, as computations can be performed on encrypted data without compromising privacy.

*3.3. Zero-knowledge proof (ZKP)*

In the ZKP scheme, there exist two essential entities, a prover $\mathcal{P}$ and a verifier $\mathcal{V}$. ZKP is a cryptographic method in which the prover $\mathcal{P}$ can convince the verifier $\mathcal{V}$ that a statement is true without revealing any information other than the truthfulness of the statement itself [35].

A ZKP scheme comprises a proof generation function *Generation* () and a proof verification function *Verification* (). The input to the proof generation function is a statement or a question, denoted as $Q$. The prover $\mathcal{P}$ executes the *Generation* ($Q$) function, which generates a ZKP $P$. Upon performing the *Verification* (P) function, the verifier $\mathcal{V}$ will be able to determine the output as either True or False.

**4. Protocol**

This section outlines our proposed privacy-preserving solution for utilizing private data in DeFi protocols. We begin by introducing the privacy problem we aim to solve, followed by the introduction of the system participants. We then present a detailed process and provide security proof for our system.

## 4.1. Problem statement

This subsection formally describes the privacy problem between users and DeFi protocols in the financial data security scenario in DeFi.

**Definition 1** (User). *Let U represent an entity seeking access to DeFi protocol services by verifying its asset information.*

User $U$ possesses the following properties:
1. Asset Ownership: User $U$ owns assets that are distributed across different institutions, denoted by $A_1, A_2, ..., A_n$, where each $A_i$ represents the number of assets held. The sum of assets $A_1, A_2, ..., A_n$ is represented by $result_U$.
2. On-chain Address: User $U$ possesses multiple addresses on the blockchain, each associated with different blockchain assets. However, when applying for DeFi protocol services, User $U$ utilizes a specific on-chain address, denoted as $addr_U$.

**Definition 2** (DeFi protocol). *A DeFi protocol D in decentralized finance can be mathematically represented as an application characterized by the following components:*
1. Operator: The DeFi protocol $D$ requires an operator, who contributes information for the protocol's operation. The operator possesses an asymmetrically encrypted private key $sk_D$ and a corresponding public key $pk_D$.
2. User Information: The DeFi protocol $D$ relies on user information to provide services. This information includes the user's address $addr_U$, used for service provision, and the user's asset information, utilized for tailored services.

In practice, DeFi protocols encompass various financial services such as lending and derivatives, aiming to enhance financial systems through decentralized technologies.

However, in decentralized finance, the issue of privacy protection holds significant importance. Due to privacy concerns, users frequently express reservations about disclosing their exact asset information, particularly regarding their on-chain addresses. In our work, we define an attack model as follows:

**Definition 3** (User Privacy Attack). *An Adversary $\mathscr{A}$ can acquire User U's precise asset information $A_1, A_2, ..., A_n$ and associate it with the on-chain address $addr_U$ by compromising a single involved entity.*

While not resulting in direct asset loss on the blockchain, these attacks expose users to significant privacy breaches. The primary consequence is the linkability of transactions, enabling the inference of user information such as account balances, transaction types, and frequencies. By combining statistical data and background knowledge, adversaries can deduce the user's true identity with confidence [5]. This vulnerability substantially heightens the risks of fraudulent activities, targeted attacks, and surveillance by off-chain centralized institutions.

Moreover, the limited utilization of complete asset data by users accessing DeFi financial services adversely affects their rights and interests [36]. This limitation introduces gaps between offered loan amounts or mortgage rates and the user's actual creditworthiness. Consequently, it can detrimentally impact the overall efficiency and effectiveness of the DeFi market.

*4.2. System participants*

In response to the privacy problem in DeFi, we propose a novel solution that involves a system comprising three key stakeholders: trusted sources, a ZKPSP, and relayers. The proposed system aims to address the challenge of maintaining privacy while recording users' asset values on-chain for verification purposes. Through the collaborative efforts of these three participants, our solution, on the one hand, ensures that users' private data remains secure on the public chain. Notably, no single entity within our system can access all of a user's private data, significantly reducing privacy risks even in the event of data breaches. On the other hand, our solution prioritizes data availability while maintaining privacy, allowing DeFi protocols to enhance services using private details like user asset information.

**Definition 4** *(Trusted sources). Trusted sources, represented by $T_1, T_2, ..., T_n$, provide asset information for a user, denoted by $A_1, A_2, ..., A_n$. In our proposed system, the trusted sources should satisfy the following properties:*

1. Authenticity: Each trusted source $T_i$ will truthfully upload the owned user information $A_i$ after verifying the authenticity of the user. The verification process entails verifying both the initiating service's address and the associated sensitive information address owned by the trusted source $T_i$. An example verification method involves the trusted source encrypting a message on the blockchain using the public key of the address to be verified and requiring the user to decrypt it using the corresponding private key.
2. Independence: The actions and operations of one trusted source $T_i$ do not rely on or require the involvement or cooperation of any other trusted sources. For any $T_i, T_j$ where $i \neq j$, the actions of $T_i$ are independent of $T_j$.
3. Traceability: Each trusted source $T_i$ uploads the encrypted user information $C_i$ to the blockchain for record.

In practice, a trusted source is a trusted third-party entity that offers precise asset quantities to the user. These trusted sources can include off-chain centralized banks, on-chain exchanges, audit firms, or other recognized authorities within the system.

**Definition 5** (Relayer). *A relayer $R$ is an entity capable of performing homomorphic computation $Hom_f$, on the encrypted information $C_1, C_2, ..., C_n$ to compute the encrypted result $Cresult_U$.*

The relayer exhibits the following properties:
1. Unknowability: The relayer $R$ is unable to decrypt $C_1, C_2, ..., C_n$ and $C_{result}$, ensuring that the encrypted information remains inaccessible to it.
2. Independence: The operation of the relayer $R$ is independent of other entities and does not engage in collusion with any other subjects.

3. Traceability: The relayer $R$ uploads the accepted information and generated results to the blockchain for traceability. It possesses a pair of asymmetrically encrypted public and private keys denoted as $pk_R$ and $sk_R$ respectively.

**Definition 6** (Zero-Knowledge Proof Service Provider (ZKPSP)). *The ZKPSP, denoted by Z, is a mechanism that facilitates the generation of a ZKP $\mathscr{P}$ to demonstrate that the user's asset belongs to a specific range while preserving the confidentiality of the actual result representing the user's overall asset information. Other entities can employ the algorithm $\mathscr{V}$ to verify the validity of the proof $\mathscr{P}$.*

ZKPSP $Z$ also maintains the properties of independence and traceability and possesses a pair of asymmetrically encrypted public and private keys, denoted as $pk_Z$ and $sk_Z$, respectively.

*4.3. Workflow*

In this subsection, we describe the workflow of our proposed financial privacy system for DeFi. The process bifurcates into two phases: initially, from the user's perspective, where they submit private asset data to the blockchain; subsequently, from the DeFi protocol's stance, which retrieves secure asset information from the blockchain via interactions with our privacy system. Those two processes ensure privacy preservation and data availability aligned with the specific requirements of DeFi.

4.3.1 Uploading asset information

We present the workflow for uploading asset information onto the blockchain network in a privacy-preserving manner (refer to Figure 1). Our approach emphasizes the encryption of asset information to ensure that it remains hidden from all system participants (refer to Figure 2). This method protects sensitive data from potential exposure while maintaining the traceability of transactions.

• **Request Initialization.** The initial phase of uploading asset information commences with the request initialization. A user, referred to as Alice, initiates the request to trusted sources in order to obtain asset information. To safeguard her privacy, Alice employs the public key $pk_D$ provided by the DeFi protocol she intends to engage with, to encrypt her blockchain address $addr_U$. This encryption process produces an encrypted address denoted as $Caddr_U$.

• **Encryption Preparation.** After encrypting her address, Alice sends $Caddr_U$ to some independent trusted sources, who encrypt Alice's original asset data $A_1, A_2, ..., A_n$ with the ZKPSP's public key $pk_Z$, resulting in encrypted asset information sets $C_1, C_2, ..., C_n$. These encrypted asset sets $C_1, C_2, ..., C_n$ are further encrypted with the relayer's public key $pk_R$ to obtain $E_{C_1}, E_{C_2}, ..., E_{C_n}$. The trusted sources then upload the encrypted asset information along with $Caddr_U$ to the relayer via the blockchain system. This workflow ensures that the sensitive asset information is encrypted at each stage and remains hidden from potential attackers on the blockchain.

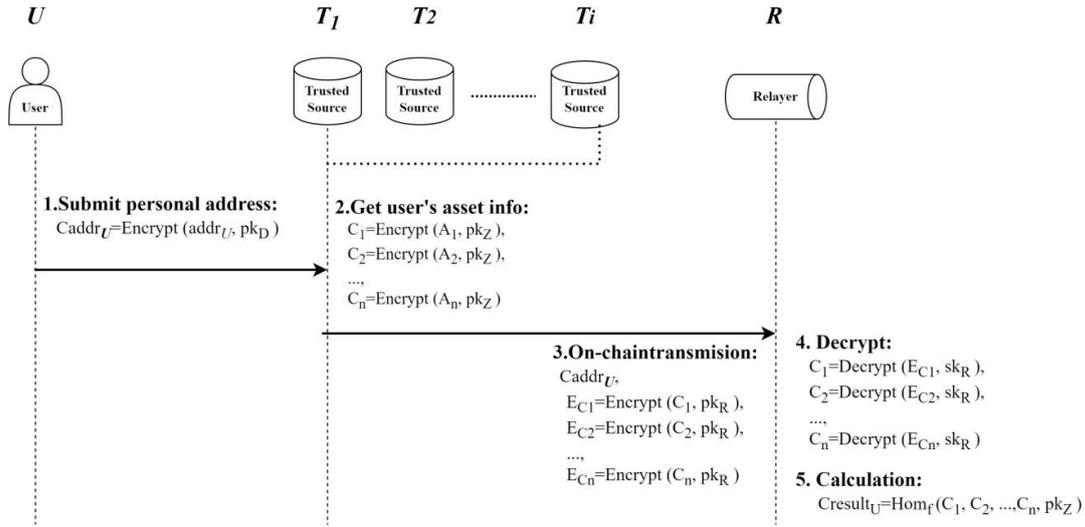

**Figure 1.** The workflow for uploading asset information onto the blockchain network.

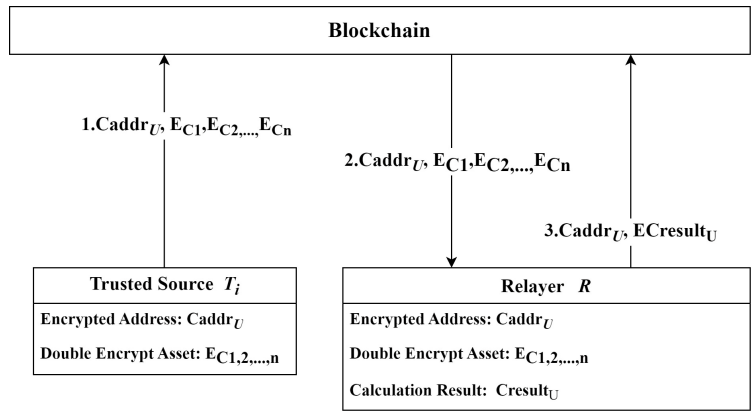

**Figure 2.** The upload of asset information is a read-and-write operation on the blockchain.

• **Homomorphic Computation**. After Alice's asset information is encrypted and uploaded to the blockchain, the relayer is responsible for performing computations on the ciphertext to ensure that the asset information remains private. Specifically, the relayer retrieves the encrypted asset information and decrypts it using its private key, $sk_R$, to obtain $C_1$, $C_2$, ..., $C_n$. The relayer then applies a homomorphic computation function, denoted by $Hom_f$, to the encrypted asset information, resulting in $Cresult_U$. Finally, the relayer uploads $Cresult_U$ and $Caddr_U$ to the blockchain, ensuring that the asset information remains confidential.

4.3.2 Asset verification

After uploading the asset information onto the blockchain, DeFi protocols can evaluate the data to determine whether it meets their requirements for updating financial services. However, the asset verification process is not automated and requires user interaction with the DeFi protocol to initiate verification with the help of the DeFi protocol operator. Figure 3 provides a visual representation of the asset verification workflow, while Figure 4 depicts the information exchange between stakeholders and the blockchain framework during this stage.

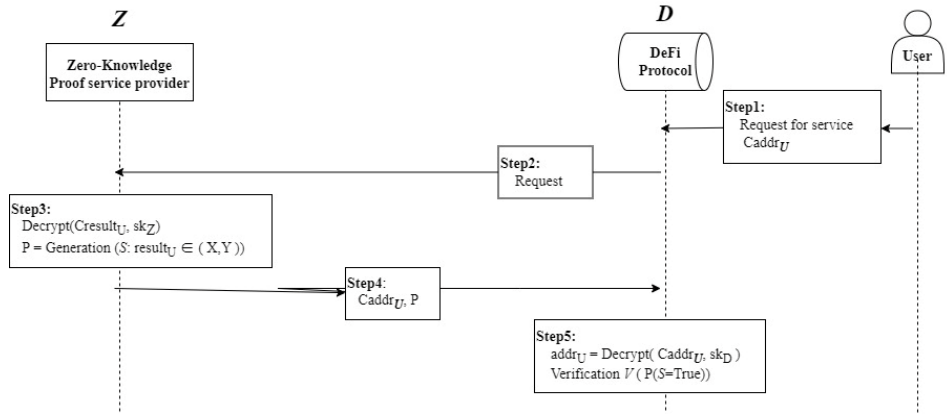

**Figure 3.** Workflow of Asset Verification.

- **Proof Request.** The asset verification phase commences with the DeFi protocol sending a proof request to the financial privacy system. The request contains the user's encrypted address, denoted by $Caddr_U$ from the user $U$ who wants DeFi protocol service, and the statements $S$. The statement $S$ asserts that ZKPSP $Z$ possesses knowledge of user $U$'s aggregate assets, represented as $result_U$. The specific content of the statement $S$ is determined by the requirements of the DeFi protocol. For instance, in this scenario, let's assume that the DeFi protocol $D$ offers different services based on users' asset levels. When Alice, as user $U$, applies for services, the DeFi protocol $D$ needs to determine the appropriate asset level associated with the requested service. To accomplish this, the DeFi protocol divides users into two intervals: $(W, X]$ and $(X, Y]$. Consequently, two statements $S$ are sent to ZKPSP $Z$: "Alice's total asset $result_U$ belongs to the interval $(W, X]$" and "Alice's total asset $result_U$ belongs to the interval $(X, Y]$".

- **ZKP Generation**. Upon receiving the asset verification request from the DeFi protocol, ZKPSP $Z$ initiates a ZKP procedure. ZKPSP $Z$ acts as the prover in the ZKP scheme and leverages its private key $sk_Z$ to decrypt $Cresult_U$, thus obtaining the computation result of the asset information. Subsequently, ZKPSP $Z$ generates a ZKP based on asset information $Cresult_U$, which is recorded as $\mathscr{P}$. This $\mathscr{P}$ is based on the statements: Alice's total asset $result_U$ belongs to the interval $(X, Y]$ is True while Alice's total asset $result_U$ belongs to the interval $(W, X]$ is False. ZKPSP $Z$ then shares the ZKP $\mathscr{P}$ and the encrypted user address $Caddr_U$ with the DeFi protocol $D$ via the blockchain system. This approach guarantees that the DeFi protocol $D$ can verify Alice's asset information without knowing the precise asset value, thereby preserving user privacy.

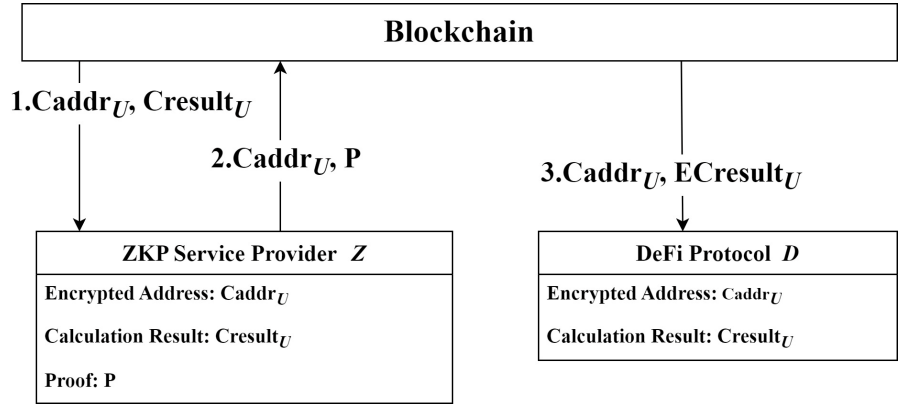

**Figure 4.** Information is both read and written onto the blockchain during the asset verification phase.

- **ZKP Verification and Financial Services Update.** Upon receiving the ZKP $\mathscr{P}$ from ZKPSP $Z$, the DeFi protocol operator $D$ performs a verification step using the algorithm $\mathscr{V}$ and can ensure that Alice's $result_U$ belongs to the interval $(X, Y)$. The DeFi protocol operator $D$ decrypts Alice's encrypted address using its own private key $sk_D$ to identify Alice and updates the financial services provided to Alice accordingly. The DeFi protocol is capable of utilizing the user's financial information without revealing the user's privacy on the blockchain. This process ensures that the DeFi protocol can execute its functions while also preserving the confidentiality of user information.

*4.4. Security proof*

After introducing our proposed financial privacy system for DeFi, our validation process will encompass two critical aspects. First, we will evaluate its effectiveness in addressing the attack model as outlined in Section 3. Second, we will assess the protocol's robustness and security against potential malicious users.

4.4.1 Attack model

Our analysis demonstrates that none of the participants in the system can obtain precise financial information $A_1, A_2, ..., A_n$ of a particular DeFi user $U$.

**Lemma 1.** An Adversary $\mathscr{H}$ can not acquire User $U$'s precise asset information $A_1, A_2, ..., A_n$ and associate it with the on-chain address $addr_U$ by compromising **trusted sources** $T_1, T_2, ..., T_n$.

*Proof.* Trusted sources $T_1, T_2, ..., T_n$ hold information regarding users' assets $A_1, A_2, ..., A_n$. but as they do not have the DeFi protocol operator's public key $pk_D$, as referred in the section 3.1., they can not encrypt the user address $Caddr_U$ stored on the blockchain. Similarly, they can not encrypt $Cresult_U$ as a lack of ZKPSP's private key $sk_Z$. Therefore, if the Adversary $\mathscr{H}$ attacks a trusted source $T_i$, all the Adversary $\mathscr{H}$ can know is the related users' assets $A_i$.

**Lemma 2.** An Adversary $\mathcal{A}$ can not acquire User $U$'s precise asset information $A_1, A_2, ..., A_n$ and associate it with the on-chain address $addr_U$ by compromising the **relayer $R$**.

*Proof.* As for the relayer $R$, it receives $E_{C_1}, E_{C_2}, ..., E_{C_n}$ from the trusted sources and decrypts them to $C_1, C_2, ..., C_n$ and $Cresult_U$ as the ciphertexts of homomorphic encryption. As the relayer $R$ does not have the DeFi protocol operator's public key $pk_D$, as referred in the section 3.2., it can not encrypt user address $Caddr_U$ stored on the blockchain. Similarly, all of these ciphertexts $C_1, C_2, ..., C_n$ and $Cresult_U$ are encrypted using ZKPSP's public key $pk_Z$, which can only be decrypted using ZKPSP's private key $sk_Z$. As a result, the relayer is unable to know the plaintext of $C_1, C_2, ..., C_n$ and $Cresult_U$.

Therefore, if the Adversary $\mathcal{A}$ attack the relayer $R$, all the Adversary $\mathcal{A}$ can know is the ciphertexts $C_1, C_2, ..., C_n$ and $Cresult_U$.

**Lemma 3.** An Adversary $\mathcal{A}$ can not acquire User $U$'s precise asset information $A_1, A_2, ..., A_n$ and associate it with the on-chain address $addr_U$ by compromising the **ZKPSP $Z$**.

*Proof.* As for the ZKPSP $Z$, it receives $Cresult_U$ and the encrypted address $Caddr_U$ from the relayer. Since $Caddr_U$ is encrypted using the DeFi protocol operator's public key $pk_D$, as referred in the section 3.1., ZKPSP is unable to decrypt it. All it can decrypt is the $Cresult_U$ as $Cresult_U$ is encrypted by its public key $pk_Z$. As a result, ZKPSP $Z$ cannot determine the exact asset value associated with a specific user $U$ on the blockchain. Therefore, if the Adversary $\mathcal{A}$ attacks the ZKPSP $Z$, all the Adversary $\mathcal{A}$ can know is an asset value result related to an unknown user.

**Lemma 4.** An Adversary $\mathcal{A}$ cannot acquire User $U$'s precise asset information $A_1, A_2, ..., A_n$ and associate it with the on-chain address $addr_U$ by compromising the **DeFi protocol**.

*Proof.* In the context of the DeFi protocol, the operator collaborates with ZKPSP $Z$ to receive a ZKP $\mathcal{P}$ based on the encrypted address $Caddr_U$. The validity of the proof is verified by utilizing the algorithm $\mathcal{V}$. Additionally, the operator possesses the private key $sk_D$ to decrypt $Caddr_U$. Consequently, if the Adversary $\mathcal{A}$ were to target the DeFi protocol, their knowledge would be restricted to User $U$'s address and the knowledge that the user's asset falls within a certain interval, without access to the precise asset value.

Therefore, we can conclude the following theorem 1:

**Theorem 1.** In our system, the Adversary $\mathcal{A}$ is unable to acquire precise asset information $A_1, A_2, ..., A_n$ and link it to the on-chain address $addr_U$ through attacks on any single involved entities.

*Proof.* By considering Lemma 1, Lemma 2, Lemma 3, and Lemma 4, it becomes evident that irrespective of the Adversary $\mathcal{A}$'s target, whether it is the trusted sources $T_1, T_2, ..., T_n$, the relayer $R$, the ZKPSP $Z$, or the DeFi protocol, its access remains restricted to only partial information concerning User $U$.

Therefore, we establish that it is impossible for any entity within the system to establish a direct association between a user's blockchain address and their financial information. This holds true regardless of the available information accessible to participants and the publicly accessible data on the blockchain. Similarly, even in scenarios where participant information may be compromised, users remain adequately protected against potential risks arising from the exposure of their financial privacy. Consequently, the privacy of users in their financial transactions is effectively ensured.

4.4.2 Malicious user

Given the potential vulnerabilities introduced by malicious entities seeking to exploit DeFi systems, the necessity for a resilient protocol is important. Our analysis indicates that malicious actions by such users yield no benefits for them.

**Theorem 2.** In our system, the malicious user $U_m$ is unable to enhance its utility through any operations within our system, and the utility of other users $U$ remains unaffected.

*Proof.* In our designed protocol, user operations are limited to the initial stages of two distinct phases: asset information upload phase and asset verification phase.

In the asset information upload phase, user $U$ interacts solely by providing address $Caddr_U$ to the trusted source $T_i$. The malicious user $U_m$ might attempt deceit by submitting an erroneous address $Caddr_{U^*}$ instead of its address $Caddr_{Um}$. However, $T_i$ has the responsibility to validate the legitimacy of the initiating user prior to conducting upload tasks. Data uploads by $T_i$ hinge on the successful verification of both the initiating address and the associated data's address held by $T_i$. Thus, should a malicious user submit $Caddr_{U^*}$, the initial authentication would fail, thereby halting further protocol proceedings.

In the asset verification phase, user $U$ is restricted to sending their address $Caddr_U$ to the DeFi protocol. The malicious user $U_m$ might send an erroneous address $Caddr_{U^*}$ from its actual address $addr_{Um}$. Nonetheless, the DeFi protocol can decrypt and discern inconsistencies between the decrypted address $addr_{U^*}$ and the requested address $addr_{Um}$. When discrepancies arise between $addr_{U^*}$ and $addr_{Um}$, the DeFi protocol denies services to $U_m$.

Considering the utility of other users $U$, only users $U^*$ associated with the wrong address $Caddr_{U^*}$ are involved in this process. As we proved above, for any malicious user $U_m$ who inputs an erroneous address $Caddr_{U^*}$, it is proved that they are denied access to services within our system. $U_m$ is precluded from expropriating any asset information associated with $Caddr_{U^*}$. Therefore, the utility of the user $U^*$ corresponding to that address $Caddr_{U^*}$ remains unaffected.

In summary, our system enhances DeFi privacy by integrating ZKP and homomorphic encryption within the blockchain structure, safeguarding user data. Homomorphic encryption facilitates the secure processing of encrypted information, preserving financial data availability. ZKP allows for data verification without disclosing sensitive content. The incorporation of RSA strengthens security, enhancing on-chain transaction protection. With the combination of ZKPSP, relayers, and trusted entities, we ensure no participant accesses full user details, thus reducing breach implications. Hence, our DeFi protocol achieves both privacy preservation and data availability tailored to DeFi's distinct needs.

## 5. Discussion

Our system, on the one hand, ensures that no participant can access the complete sensitive data of individual DeFi users, reducing breach implications. On the other hand, the combination of different functions enhances protected sensitive data accessibility. In this section, we aim to explore the possible scenarios where the proposed financial privacy system can be utilized to target the bolstering of DeFi protocol security. Specifically, we focus on two primary categories of protocols: public blockchain DeFi and cross-border financial applications. These scenarios are of significant interest to the wider community, and our proposed system offers a promising solution for addressing the privacy concerns in these domains.

### 5.1. DeFi applications on public blockchain

Risk control is a critical function in the mechanism design for financial applications, including numerous DeFi protocols such as lending protocols that rely on over-collateralization to maintain system stability. However, this approach negatively impacts capital efficiency and may result in reputable institutions engaging in DeFi lending protocols not obtaining reasonable mortgage rates, despite having better credit than many individual DeFi users. Our proposed solution aims to introduce real-world credit ratings to DeFi on public blockchain systems while preserving privacy.

By integrating credit ratings with personal assets not reserved in the DeFi protocol or specific blockchain addresses, users' collateral rates can be reduced, and collateral-free lending can be enabled. Moreover, the governance of the DeFi protocol can propose and vote on a set of criteria specifying the potential benefits based on users' credit ratings.

In addition to its application in DeFi lending protocols, data privacy protection can be utilized for non-fungible token (NFT) market design. Our credit rating system enables installment purchases of NFTs, giving users the flexibility to make purchases when they may have limited funds. By verifying credit, users can pre-purchase NFTs and make payments in installments to artists or galleries.

In summary, our proposed solution contributes to a more efficient capital utilization in DeFi on public blockchain systems by preventing over-collateralization. Moreover, assessing risk through credit ratings enhances risk control for DeFi protocols and enables installment purchases of NFTs, promoting a more diverse market.

### 5.2. Cross-border financial application

Our work aims to address the dual challenge of private information verification and privacy protection, making it a promising solution for centralized scenarios such as cross-border data transfer. In many countries, laws exist to regulate cross-border data transmission, driven by concerns for national data security and personal privacy. These regulations can hinder the growth of cross-border trade and globalization, which are heavily dependent on data exchange.

For example, in the realm of cross-border e-commerce, online information exchange and qualification verification can be costly and inadequate. Our solution enables sellers to provide encrypted information to verify the authenticity of their qualifications, thereby facilitating compliance requirements.

In cross-border borrowing markets, loan rates vary significantly across different geographical areas. Consequently, borrowers often seek to capitalize on rate differentials to minimize borrowing costs. However, the transmission of personal data across borders is strictly regulated, making it challenging for financial institutions to evaluate a borrower's creditworthiness in an area outside their own. Our system offers a feasible solution for secure information transmission in cross-border scenarios through an effective governance mechanism. The transmitted data remains confidential, and sensitive information is inaccessible to external entities.

It is important to note that cross-border activities may involve jurisdictional discrepancies, and implementing technologies such as cross-chain to address such issues is not yet fully established. While implementing cross-border borrowing services largely depends on regulatory compliance, our system provides a promising solution for data privacy protection in cross-border financial applications.

## 6. Conclusion

This paper addresses a critical issue in DeFi: the utilization of private data while ensuring privacy protection for users. By introducing three internal stakeholders and proposing a solution, our system first ensures that users' private data remains secure on the public chain. Notably, no single entity within our system can access all of a user's private data, significantly reducing privacy risks even in the event of data breaches. Secondly, our system prioritizes data availability while maintaining privacy, allowing DeFi protocols to enhance services using private details like user asset information. The proposed solution enables the utilization of multiple sensitive data while maintaining privacy in the DeFi ecosystem. This work contributes to the ongoing effort to create a more secure and privacy-preserving DeFi landscape.

**Acknowledgments**

This work is supported by The Science and Technology Development Fund, Macau SAR (File no. 0129/2022/A and no.0091/A2/2020) and the University of Macau (File no. MYRG-CRG2022-00013-IOTSC-ICI and no. SRG2022-00032-FST).

**Conflicts of interests**

The authors declared that they have no conflicts of interests.